\documentclass[%
 reprint,
 amsmath,amssymb,
 aps,
prb,
]{revtex4-2}

\usepackage{csquotes}
\usepackage{graphicx}
\usepackage{dcolumn}
\usepackage{bm}
\usepackage{hyperref}

\usepackage[colorinlistoftodos]{todonotes}

\newcommand{\Rads}{\ \mathrm{Rad\ s^{-1}}}
\usepackage{xcolor}

\begin{document}

\preprint{APS/123-QED}

\title{Transition between ground states in square anisotropic artificial colloidal ice}

\author{Leonardo G. Alanis-Cantú}
\author{Antonio Ortiz-Ambriz}
\email{aortiza@tec.mx}
\affiliation{
Tecnológico de Monterrey, Escuela de Ingeniería y Ciencias\\
Ave. Eugenio Garza Sada 2501, Monterrey, 64849, México.
}

\date{\today}

\begin{abstract}
  In Artificial Colloidal Ice (ACI), paramagnetic colloidal particles are confined in double-well traps and interact via repulsive, isotropic, magnetic dipole-dipole interactions that can be controlled by an external magnetic field. In this paper, we dynamically introduce anisotropic interactions to ACI by rotating the external magnetic field, which, in equilibrium, makes the system go from a charge-free 2-in, 2-out ice rule state, to a charged 4-in, 4-out state. 
We observe a strong dependence of the final configuration on the field's rotation rate $\omega$: at high angular velocity, the system achieves a defect free final state via a difussionless transformation from the initial ground state. However, counterintuitively, at slow rotation rates, ergodicity breaks down, trapping the system in a partially ordered metastable state.
\end{abstract}


\maketitle

%

\section{\label{sec:introduction}Introduction}

Ice-like model systems provide a versatile platform for studying geometric frustration. They consist of elongated elements with two possible orientations, arranged on the edges of a lattice so that neighboring elements meet at common vertices. Each vertex can be assigned a topological charge according to the orientations of the surrounding elements. In equilibrium, low-energy configurations generally obey a version of Pauling's ice rule, in which the local topological charge is minimized \cite{ortiz-ambrizColloquiumIceRule2019, nisoliColloquiumArtificialSpin2013}.

Among these systems, the best-known realization is artificial spin ice (ASI), composed of elongated ferromagnetic nano-islands whose magnetization is constrained by shape anisotropy to lie along their long axis, giving rise to an Ising-like degree of freedom \cite{wangArtificialSpinIce2006, skjaervoAdvancesArtificialSpin2019}. The islands interact through in-plane magnetic dipolar interactions, and the orientation of each magnetic moment relative to a vertex determines its topological charge. Advances in experimental techniques have enabled both direct imaging of the resulting magnetic configurations through magnetic force microscopy and transmission electron microscopy \cite{mayMagneticChargePropagation2021, driskoTopologicalFrustrationArtificial2017, wangRewritableArtificialMagnetic2016, gilbertDirectVisualizationMemory2015, gilbertEmergentIceRule2014, zhangCrystallitesMagneticCharges2013}, as well as real-time observations of their dynamics using x-ray magnetic circular dichroism \cite{sacconeRealspaceObservationErgodicity2023, aravaEngineeringRelaxationPathways2019, laoClassicalTopologicalOrder2018, morleyVogelFulcherTammannFreezingThermally2017, kapaklisThermalFluctuationsArtificial2014, rougemailleArtificialKagomeArrays2011}.

An alternative realization, within the broader class of particle ice systems, is \emph{artificial colloidal ice} (ACI) \cite{nisoliUnexpectedPhenomenologyParticleBased2018}. In ACI, the elementary degrees of freedom are bistable traps, each occupied by a single colloidal particle, so that the particle position within the trap defines an Ising-like variable. A key advantage of this surrogate platform is that its configurations can be directly imaged using optical microscopy. In these systems, interactions may be electrostatic \cite{libalRealizingColloidalArtificial2006} or may be induced by applying an external magnetic field to superparamagnetic particles \cite{ortiz-ambrizEngineeringFrustrationColloidal2016}.

Under appropriate conditions, ACI and ASI share the same effective vertex energetics \cite{nisoliUnexpectedPhenomenologyParticleBased2018, nisoliDumpingTopologicalCharges2014}. This correspondence can be understood by assigning to each bistable trap an effective dipole built from the real colloid and a virtual negative charge distributed between the two stable positions. In this picture, the energetics are governed by the local minimization of topological charge, so that a uniform ACI lattice also favors the 2-in/2-out ice-rule state. The analogy, however, only holds when the colloidal interactions remain isotropic and repulsive. For magnetic ACI, this condition is satisfied when the external field is perpendicular to the plane, which is the regime explored in most experimental and numerical studies to date \cite{ortiz-ambrizEngineeringFrustrationColloidal2016, rodriguez-galloDegeneracyHysteresisBidisperse2021, rodriguez-galloIceRuleBreakdown2023, libalIceRuleFragility2018, rodriguez-galloGeometricalControlTopological2023, libalInnerPhasesColloidal2018, lecunuderCompetingOrdersColloidal2019, libalQuenchedDynamicsArtificial2020, oguzTopologyRestrictsQuasidegeneracy2020, libalDopedColloidalArtificial2015}. External driving can still bias the energetic hierarchy of the vertices and promote defects or alternative ordered states \cite{ostinatoTrackingTopologicalDefects2025, libalHysteresisReturnpointMemory2012, libalDynamicControlTopological2017}, but the pairwise colloidal interactions themselves remain isotropic and repulsive, and the force acts only along the trap direction.

Two recent ACI studies move beyond this isotropic, purely repulsive regime. In Ref. \cite{baillouEmergentChargeCrystallization2026}, an in-plane rotating magnetic field generates an effective time-averaged attraction between particles, driving the system away from the ice-rule manifold and into an anti-ice-rule charge crystal with 4-in/4-out vertices on the square lattice, or more generally toward states that locally maximize charge. By contrast, Ref. \cite{alanis-cantuAnisotropicInteractionsInduce2026} considers a static field aligned with one of the lattice vectors. In that case the interactions become anisotropic, and although the ground state is again the same charge-crystal phase, the system encounters strong collective barriers that hinder relaxation and favor metastable configurations.

In this work, we study how the system evolves between the ice-rule and anti-ice-rule states under a continuous rotation of the external magnetic field from the out-of-plane direction toward the particle plane. Using simulations over a range of field strengths $B$ and angular frequencies $\omega$, we identify two distinct dynamical regimes. For fast rotations, the system undergoes a collective and highly coordinated reconfiguration reminiscent of a diffusionless transformation, reaching the antiferromagnetic ground state without large-scale particle diffusion. For slow rotations, by contrast, the system does not fully equilibrate and instead becomes trapped in long-lived metastable configurations.


\section{\label{sec:methods}Simulation}

\begin{figure}
  \begin{center}
    \includegraphics{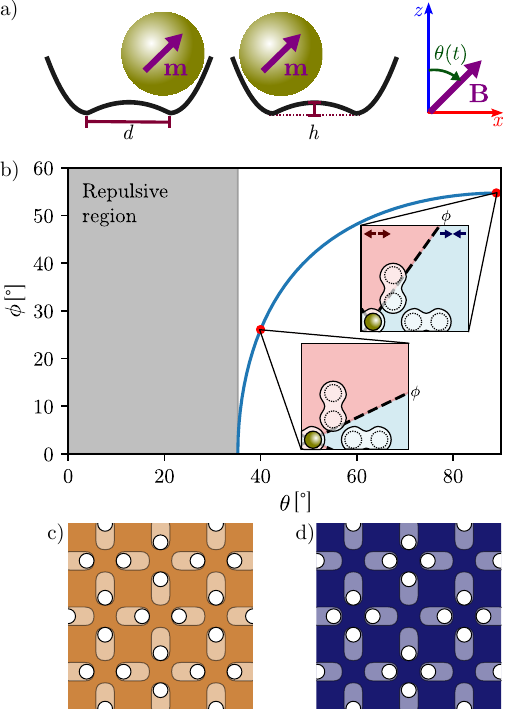}
  \end{center}
  \caption{Simulation setup and dipolar interactions. (a) Colloidal particles are confined in double-well traps with an externally rotating magnetic field that induces dipole-dipole interactions. The field rotates starting from the $z$-axis until it reaches the $x$-axis with a rate $\omega=\mathrm{d}\theta/\mathrm{d}t$, where $\theta$ is the polar angle. (b) Angle separating attractive and repulsive interactions in the $xy$ plane, i.e. the magic angle, as a function of the rotation angle $\theta$. The inset images show how the rotation of the field starts by attracting only the collinear particle, and eventually the attractive region includes the particle in the perpendicular trap. The gray region indicates where all interactions are repulsive. (c) A region of vertices in the 2-in/2-out ice rule and (d) vertices in the 4-in/4-out anti-ice rule.}\label{fig:system}
\end{figure}

We simulate paramagnetic colloidal particles of radius $r = 5\,\mu\textrm{m}$, magnetic susceptibility $\chi = 0.0576$, and density $\rho = 1\,\textrm{g}/\textrm{cm}^3$, confined in double-well traps characterized by a separation $d = 10\,\mu\textrm{m}$ and trap height $h = 4\,\textrm{pN}\cdot\textrm{nm}$ (see Fig.~\ref{fig:system}a). The traps are placed along the edges of a $10 \times 10$ square lattice with lattice constant $a = 30\,\mu\textrm{m}$, following a configuration that matches what has been experimentally realized~\cite{ortiz-ambrizEngineeringFrustrationColloidal2016}.

Time evolution is done using overdamped molecular dynamics simulations. At each time step, we integrate the equations of motion:
\begin{equation}
  \gamma \frac{d\mathbf{r}_i}{dt} = \mathbf{F}_i^{\textrm{dd}} + \mathbf{F}_i^{\textrm{traps}} + \bm{\eta}_i,
  \label{eq:eom}
\end{equation}
where the drag coefficient $\gamma = k_B T/\xi$ is defined from the diffusion constant $\xi = 0.125\,\mu\textrm{m}^2/\textrm{s}$, and $\bm{\eta}_i$ is a stochastic thermal noise term with zero mean and correlation $\langle \bm{\eta}_i(t)\cdot\bm{\eta}_j(t') \rangle = 4k_B T \gamma \delta_{ij}\delta(t - t')$, assuming room temperature ($T = 300\,\textrm{K}$).
Traps are modeled as bistable potentials following what is described in Refs.~\cite{rodriguez-galloTopologicalBoundaryConstraints2021, alanis-cantuAnisotropicInteractionsInduce2026}, 
\begin{eqnarray}
    \mathbf{F}_i^{\textrm{trap}} &=& -k_\textrm{trap}r_\perp \hat{\mathbf{e}}_\perp \nonumber \\ 
    &-& \hat{\mathbf{e}}_\parallel 
    \begin{cases}
        -k_{\textrm{hill}}r_\parallel & |r_\parallel|\leq d/2,\\
        k_{\textrm{trap}}(|r_\parallel| - d/2)\textrm{sgn}(r_\parallel) & |r_\parallel| > d/2,
    \end{cases}
\end{eqnarray}
with $k_\textrm{trap} = 0.1\,\textrm{pN}/\textrm{nm}$ and $k_\textrm{hill} = 8h/d^2$. In our simulations all hills have the same height, with no disorder. 

An external magnetic field $\mathbf{B}$ induces a magnetic dipole moment on each particle, making particles interact through magnetic dipole–dipole interactions
\begin{equation}
  U_{ij} = -\frac{\mu_0 m^2}{4\pi r_{ij}^3} \left[ 3(\hat{\mathbf{B}} \cdot \hat{\mathbf{r}}_{ij})^2 - 1 \right],
  \label{eq:dipolepotential}
\end{equation}
where $m = V\chi B/\mu_0$, $V$ is the particle volume, $\chi$ is the susceptibility, and $\mu_0$ is the magnetic constant. The unit vector $\hat{\mathbf{B}}$ points in the direction of the field, and $\hat{\mathbf{r}}_{ij}$ is the unit vector along the separation $\mathbf{r}_{ij} = \mathbf{r}_i - \mathbf{r}_j$. We assume that all dipoles have equal magnitude, and neglect many-body effects. The system is initialized in a 2-in/2-out configuration with the external field aligned along the $z$-axis (orthogonal to the particle's plane), then the field is rotated on the $xz$-plane according to the following protocol:
\begin{equation}
  \mathbf{B}(t) = B \left[ \sin(\omega t)\mathbf{\hat{x}} + \cos(\omega t)\mathbf{\hat{z}} \right], \quad 0 \leq \omega t \leq \frac{\pi}{2}.
  \label{eq:fieldprofile}
\end{equation}
All simulations were run with a time step of $0.1\,\textrm{ms}$ at constant field magnitude $B$ and angular frequency $\omega$, starting from a $2$-in/$2$-out configuration for convenience; however, this configuration can be achieved by a linear magnetic quench in the $\hat{z}$ direction.

The attractive and repulsive regions are illustrated in Fig. \ref{fig:system}b). 
When the magnetic field is homogeneous and oriented orthogonal to the particle's plane, the induced dipolar interactions are repulsive and isotropic. 
The interplay between these repulsive interactions and the constraint of topological charge conservation eventually drives the system into a 2-in/2-out ice-rule state, shown in Fig. \ref{fig:system}c).

On the other hand, when the field has a component along the $x$-axis, the interactions become both attractive and repulsive, depending on the angle $\phi$ between the line connecting the particles and the horizontal axis. The cutoff angle at which interactions change from repulsive to attractive is known as the \emph{magic angle}, which for a fully horizontal field is $\phi_m = 54.73^\circ$. 

For a dynamic field parametrized by $\theta=\omega t$, the \emph{magic angle} depends on the polar angle $\theta$. Defining $\phi$ as the azimuthal angle of the separation vector $\mathbf{\hat{r}}_{ij}$, the inner product in Eq. \ref{eq:dipolepotential} becomes $\mathbf{\hat{m}}\cdot \mathbf{\hat{r}}_{ij} = \sin(\theta)\cos(\phi)$. Therefore, the angle $\phi$ in the $xy$ plane at which interactions change from repulsive to attractive at a given field angle $\theta$ is the solution to the equation:
\begin{equation}
  \sin(\theta)\cos(\phi) = \displaystyle\frac{1}{\sqrt{3}}.
  \label{eq:magicangle}
\end{equation}

For any rotation angle $\theta<35.26^\circ$, the solution is complex, which means that the interactions remain purely repulsive and no magic angle exists. For $\theta > 35.26^\circ$, attractive regions with point symmetry appear and then grow monotonically until the magic angle saturates at $54.73^\circ$ (see Fig. \ref{fig:system}b)). We refer to this threshold as $\theta_\mathrm{th}$. Just above this threshold, the attractive region first reaches the neighboring horizontal trap. At larger values of $\theta$, the boundary between attractive and repulsive interactions cuts through the neighboring vertical trap as well. Because the attractive contribution in Eq. \ref{eq:dipolepotential} is larger than the repulsive one, a configuration with two diagonal particles close to each other becomes energetically favored over one with one particle pointing into the vertex and the other pointing out. The resulting lowest-energy state was shown in Ref. \cite{alanis-cantuAnisotropicInteractionsInduce2026} to be the 4-in/4-out anti-ice-rule state, illustrated in Fig. \ref{fig:system}d).

\section{Time Evolution and Ordering}

\begin{figure}
  \begin{center}
    \includegraphics{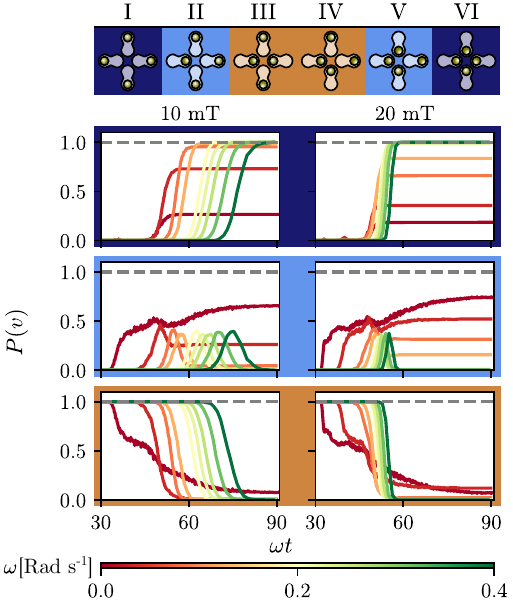}
  \end{center}
  \caption{Probability $P(v)$ of finding any given vertex type $v$ as a function of rotation angle $\omega t$ for several angular frequencies $\omega$ and two magnetic field values $B$ (columns). The vertex types are grouped by topological charge (rows), with the background indicating the color scheme shown in the top panel. The top panel (dark blue) represents the highly charged vertices of type I and VI, the middle panel represents the type II and V vertices, and the bottom panel shows the fraction of ice-rule vertices of type III and IV. 
  }\label{fig:verticescounts}
\end{figure}

We begin our analysis by examining the system's time evolution, shown in Fig. \ref{fig:verticescounts}. We estimate the probability of finding any given vertex type $P(v)$ by computing the vertex fractions at each point in time for different magnetic field magnitudes $B$ and angular frequencies $\omega=\pi/2\tau$. Here $\tau$ is the total time of rotation. In particular, we focus on two field strengths, 10~mT (low) and 20~mT (high), shown as columns in Fig.~\ref{fig:verticescounts}. Vertex types are grouped in rows by topological charge $|q|$ with the color scheme shown in the top panel.

For both magnetic field strengths, neutral vertices eventually vanish from the system as attractive interactions make them energetically unfavorable compared to charged vertices. The time at which these vertices disappear depends on the field strength, shifting to smaller angles for stronger fields. Furthermore, neutral vertices vanish at significantly smaller angles under slower rotation. Regardless of the angular frequency, substantial vertex rearrangements cannot occur before the appearance of attractive interactions. This sets a lower bound below which we do not expect vertex dynamics to significantly evolve. However, we observe some type-III vertex dynamics just below the threshold angle $\theta_\textrm{th}$, particularly at slow frequencies, indicating that even without an attractive region, the anisotropy of the interaction is enough for particles to still move to the other side of their trap and explore phase space. In other words, the first spin flips cannot be attributed to attractive interactions but to anisotropic repulsive interactions and thermal motion.
The second row of Fig. \ref{fig:verticescounts} shows peaks in the populations of the $q = \pm 2$ vertices, corresponding to transient metastable states that precede the formation of the $q = \pm 4$ ground-state vertices. These peaks arise only when the population of ground-state vertices begins to increase, and their width provides a measure of the stability of these intermediate states: the narrower the peak, the faster the decay.

Focusing on the role of angular frequency, we observe that under fast rotations, the system maximizes the fraction of $q = \pm 4$ vertices, indicating that it reaches the anti ice-rule ground state. In contrast, under slow rotation, the system settles into a mixed configuration of charged vertices and fails to reach the ground state. These mixed configurations are composed mainly of vertices with charge $q=\pm 2$ and $q=\pm 4$. This behavior is consistent across different magnetic field strengths, but the specific value of the transition angle depends on the field.

\begin{figure}
  \begin{center}
    \includegraphics{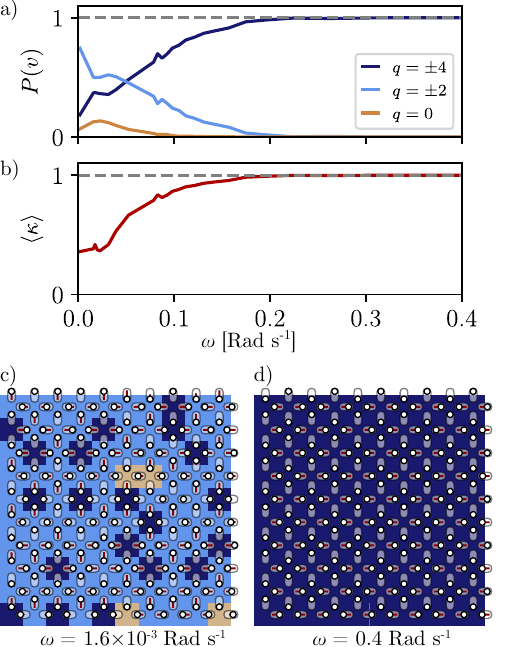}
  \end{center}
  \caption{Frequency dependence of the final configurations. (a) Probability $P(v)$ of finding a vertex $v$ with charge $q$ in the last frame of a simulation as a function of angular frequency $\omega$. (b) Charge order parameter $\kappa$ for the corresponding frames. The parameter $\kappa$ is maximized in the antiferromagnetic GS and minimized in 2-in/2-out configurations.
(c and d) Final states of a fast rotation (c) and a slow rotation (d). The background color indicates the charge of the vertex, and the red line inside the trap shows the trajectory followed by the particle.}\label{fig:wdepndence}
\end{figure}

To clarify the dependence on the angular frequency, in Fig. \ref{fig:wdepndence}a we plot the vertex fraction in the last frame of each simulation against $\omega$ for $B=20$ mT. 
The figure shows that slow rotations produce states dominated by $q=\pm2$ vertices, with relatively few $q=\pm4$, as exemplified by the configuration shown in Fig. \ref{fig:wdepndence}c). By contrast, for $\omega \gtrsim0.2 \Rads$ the system crystallizes into the anti-ice-rule ground state, shown in Fig. \ref{fig:wdepndence}d). In Fig. \ref{fig:wdepndence}b, we also show the corresponding charge order parameter, defined as
\begin{equation}
  \kappa = \dfrac{1}{4N}\displaystyle\sum_{i,j}(-1)^{i+j}q_{ij},
  \label{eq:kappa}
\end{equation}
where $N$ is the number of vertices, and $q_{ij}$ is the topological charge at vertex $(i,j)$ in the lattice. The absolute value $|\kappa|$ is maximized by a fully charged, antiferromagnetically ordered lattice, i.e. the 4-in/4-out anti-ice-rule state, and minimized by states that minimize local topological charge.

Simulations with $\omega\gtrsim 0.2 \Rads$ reach the GS, in accordance with what was observed in Fig. \ref{fig:verticescounts}. Most notably, however, slow rotations do not lead the system into an ordered state: the slowest simulation, with $\omega \approx 1.6\times10^{-3}\Rads$, reaches only $\kappa \approx 0.36$, indicating partial anti-ice-rule ordering with many frozen defects. This behavior contrasts with the usual expectation for quenched systems, where a slower quench typically produces coarser domains and cleaner ordered states \cite{libalQuenchedDynamicsArtificial2020, alanis-cantuAnisotropicInteractionsInduce2026}. It therefore points to a nontrivial timescale-dependent response arising from the continuously changing potential landscape during slow rotations.

\section{Collective Synchronized Motion}

To understand this, we now focus on the specific pathway by which the system reaches the GS. Starting from the initial 2-in/2-out ice-rule configuration, the transition into the 4-in/4-out anti ice-rule GS requires $N/2$ flips, where $N$ is the number of vertices. This corresponds to flipping exactly half of the total particles in the system. There is a clear path that connects these two states, which consists of flipping either all the vertical or all the horizontal spins, resulting in a 4-in/4-out state or its complementary configuration. This behavior is seen in Fig. \ref{fig:wdepndence}d), where the red lines inside the trap show the trajectories that the particles followed to reach the final ordered state. 

\begin{figure}
  \begin{center}
    \includegraphics{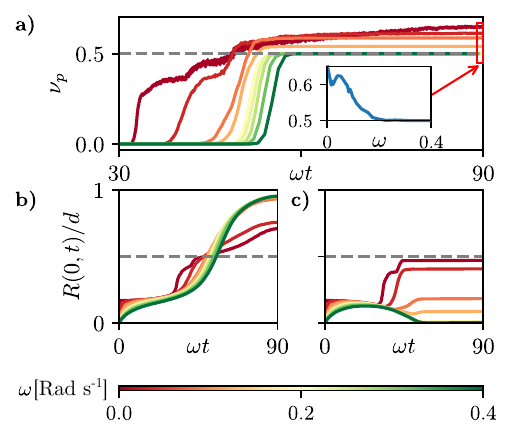}
  \end{center}
  \caption{In-trap motion of particles. (a) Cumulative fraction of particles $\nu_p$ that have crossed the central hill at time $t$. (b)-(c) Average one-dimensional in-trap travel distance, in units of trap separation, for (b) horizontal and (c) vertical traps.}
  \label{fig:intrapmotion}
\end{figure}

In Fig.~\ref{fig:intrapmotion}a we show the fraction of particles $\nu_p$ that have crossed to the other side of their trap as the field is rotated. Here we consider a particle to have crossed to the other side if it crosses the center of the hill in the direction opposite to its initial position. If a particle fluctuates rapidly around the center of the trap, it produces fluctuations in the average value of $\nu_p$, but it does not keep adding to the fraction of crossed particles. 
 
In slow rotations, particles cross the hill at different times in batches with large fluctuations, indicating that they tend to oscillate around the central hill before reaching an equilibrium position. The final value of $\nu_p$ eventually saturates well above $1/2$, which means that more than half of the particles are crossing, and the system takes a suboptimal path to the GS, which ends in a metastable state. 
In contrast, for fast rotations, $\nu_p$ increases monotonically over a brief period of time and saturates at $\nu_p = 1/2$. The slope of the curve suggests that most of the motion occurs almost simultaneously. The inset displays the last frame of each simulation as a function of $\omega$, showing behavior consistent with Fig. \ref{fig:wdepndence}.

The pathway from the ice-rule to the anti-ice-rule state involves flipping either all horizontal or all vertical traps. Since both options lead to complementary states, reaching the GS requires that only one of these two subsets flips.
In Fig. \ref{fig:wdepndence} b) and c), we examine separately the motion of horizontal and vertical particles. For this, we compute the one-dimensional trajectory of the particles by calculating the distance traveled at time $t$ relative to the initial position along the trap direction:
\begin{equation}
  R(0,t) = \langle |\mathbf{r}^{(i)}_\parallel(t) - \mathbf{r}^{(i)}_\parallel(0)| \rangle_{i,\textrm{ens}},
  \label{eq:displacemnt}
\end{equation}
where $\mathbf{r}_\parallel$ is the component parallel to the trap's direction, and $\langle \, \cdot \, \rangle_{i,\textrm{ens}}$ is the average over all particles and realizations. For the most part, particles in horizontal traps cross the hill, with the only exception being very slow fields, where still more than half the particles cross their trap. The biggest difference appears in the vertical traps shown in Fig.~\ref{fig:intrapmotion}c). For fast rotations, particles begin moving towards the other side but soon go back. However, as the rotation rate decreases, some particles continue this movement and end up crossing the hill. For very slow rotations, around half of the particles end up on the other side of the trap. 
This confirms that, at high frequencies, all particles in horizontal traps undergo a flip, and none of the vertical traps, explaining the observed $1/2$ saturation in $\nu_p$ and validating the theoretical transition pathway from the 2-in/2-out to the 4-in/4-out state.

\begin{figure}
  \begin{center}
    \includegraphics{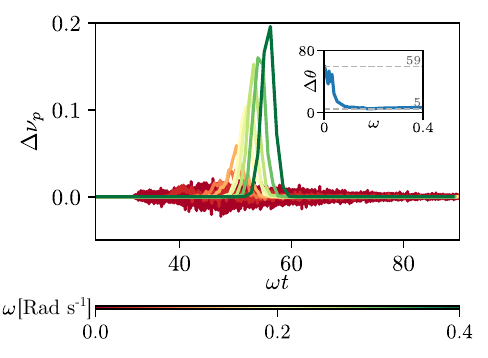}
  \end{center}
  \caption{Number of particles $\Delta \nu_p$ that cross the hill at a given point in time for different rotation rates. The inset shows the size of the region in the space of $\theta$ where flips occur, as a function of the rotation rate.}
  \label{fig:dcounts}
\end{figure}

In Fig.~\ref{fig:dcounts}, we plot the fraction of particles that crossed the hill at each time point, $\Delta \nu_p$, computed as the difference between consecutive points in Fig.~\ref{fig:intrapmotion}a. The rapid fluctuations in the slow regime indicate that particles oscillate around the central hill before finding their final state. These fluctuations span a large duration due to the scaling of time with $\Delta t = \Delta \theta / \omega$. In contrast, fast rotations yield sharp, fluctuation-free transitions, indicating a coordinated, diffusionless transformation.

At low rotation frequencies, we observe fluctuations around zero throughout the simulation, indicating persistent motion around the central hill. This suggests a continuously evolving energetic landscape that inhibits the system from reaching equilibrium. These oscillations begin at $\omega t \approx 31.77^\circ$, just below the critical angle for anisotropic interactions, $\theta = 35.26^\circ$. Therefore, the initial flips are not due to attractive interactions, but rather to a combination of weak anisotropic repulsion and thermal fluctuations, as the central hill's energy barrier is on the order of $h \sim k_B T$.

The inset in Fig.~\ref{fig:dcounts} shows the range where flips occur $\Delta \theta$, extracted from the flip-rate signal in the main panel. For each driving frequency we compute the finite difference $\Delta \nu_p$, construct a non-negative activity envelope from $|\Delta \nu_p|$ using a short moving-average smoothing, and define $\Delta \theta$ as the full angular span over which this envelope remains above a small fixed fraction of its maximum. In this way, $\Delta \theta$ measures the total window of significant activity: narrow, well-defined peaks at high frequency give small $\Delta \theta$, whereas broader and more fluctuating responses at low frequency produce larger values, consistent with prior observations. 
For reference, in the slowest cases, flips occur over $\sim654$~s, whereas in the fastest case shown, the transition completes in about $1.3$~s, again indicating a synchronized, diffusionless transition

We thus identify distinct dynamical regimes depending on the rotation speed. At high frequencies, the system undergoes synchronized, diffusionless motion that drives it into the anisotropic 4-in/4-out ground state. At low frequencies, particles fluctuate around the central hill, revealing a complex, evolving energy landscape that prevents full relaxation into equilibrium, a phenomenon likely driven by the competition between the time required for a particle to jump from one side to the other, and the timescale over which the field moves from vertical to horizontal. 

\section{Dependence on the interaction strength}

\begin{figure}
  \begin{center}
    \includegraphics{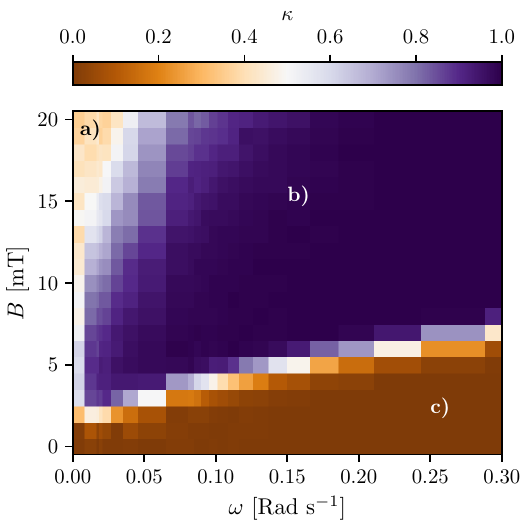}
  \end{center}
  \caption{Order parameter $\kappa$ as a function of field and angular velocity. The purple region (b) represents the parameters for which the system falls into the ordered anti-ice-rule state, and the light orange region (a) to the left shows where the system falls into a partially-ordered state. In the quiescent region (c), $\kappa \approx 0$ because the rotation is fast enough that the initial conditions are still retained when the system is observed.}\label{fig:phasediagram}
\end{figure}

Using the charge order parameter $\kappa$, we classified the system into three distinct regimes as a function of the field magnitude $B$ and angular frequency $\omega$, as shown in Fig.~\ref{fig:phasediagram}. The first region corresponds to the anti ice-rule ground state, characterized by maximized topological charge at each vertex (Fig.~\ref{fig:wdepndence}d). This state predominantly appears at high field strengths and relatively high frequencies.

The second region is a quiescent regime in which the system remains in its initial 2-in/2-out configuration (Fig.~\ref{fig:phasediagram}). This occurs either when the rotation is too rapid for the particles to respond or when the interactions are too weak to induce motion. In the high-frequency cases, this regime appears because the simulations stop at the end of the rotation, and the subsequent evolution at constant field is not included in the protocol.

The third region consists of partially ordered configurations and emerges at very low rotation frequencies. In this regime, vertices tend to acquire high topological charge, but the disorder means the overall charge is not maximized, as shown in Fig.~\ref{fig:wdepndence}c). Notably, this region broadens with increasing field strength, indicating that under slow rotations the evolving energy landscape prevents the system from fully relaxing into the GS, at least within the accessible simulation times. We expect this behavior to ultimately lead to a broken-ergodicity state in the strong-interaction limit $B\to\infty$, similar to the observations in Ref. \cite{alanis-cantuAnisotropicInteractionsInduce2026}.

\section{\label{sec:conclusion}Conclusions}

In this work, we explored the transition from the isotropic 2-in/2-out ground state to the 4-in/4-out state under a rotating external field, which exposes the system to a time-dependent energy landscape with mixed interactions. We found that the final state depends strongly on the angular frequency $\omega$: unexpectedly, fast rotations ($\omega > 0.2 \Rads$) drive the system to the GS through a diffusionless transformation characterized by synchronized motion and consistent with the optimal pathway. By contrast, slow rotations leave the system in a configuration with only partial antiferromagnetic order and apparently broken ergodicity.

These results are unexpected considering previous research \cite{libalQuenchedDynamicsArtificial2020, alanis-cantuAnisotropicInteractionsInduce2026}, where faster quenches are generally associated with larger defect densities. In our case, however, faster rotation leads to fewer defects, suggesting that this type of driving protocol has a distinct dynamical complexity and is fundamentally different from other magnetic quenches. The pronounced dependence of the final state on the driving protocol is reminiscent of kinetically arrested systems, where the accessible configuration depends on the history of the control parameter rather than solely on its final value.

We mapped three regions in the parameter space $(\omega, B)$: the AF ground state, a quiescent regime, and a partially ordered state that emerges only at low frequencies and expands as the field increases. Notably, horizontal cuts across this diagram, together with the strong-interaction limit $B\to\infty$, suggest that only a subset of the tunable parameters can drive the system into the AF state within the present protocol and accessible MD timescales. This raises the question of whether the partially ordered regime reflects ordinary metastability, kinetic arrest, genuine glass-like dynamics, or a more general form of protocol-dependent nonequilibrium ordering. Future work could address this question using enhanced sampling or machine-learning-assisted equilibrium sampling methods, such as Boltzmann Generators \cite{noeBoltzmannGeneratorsSampling2019}, to explore this region of parameter space more systematically.

\section*{Acknowledgements}

We express our gratitude to Pietro Tierno and Mallar Ray for their encouragement and their suggestions. L.G.A.-C. acknowledges financial support from SECIHTI through its graduate scholarship program. This project received support from Tecnologico de Monterrey Challenge-Based Research Funding Program through grant CI-EIC-CLS-S-116. 

\bibliography{Diffusionless.bib}

\end{document}